\documentclass[10pt,letterpaper,twocolumn]{article}

\usepackage{ol2}

\begin{document}
\twocolumn[

\title{
Femtosecond filamentation in air and higher-order nonlinearities
}

\author{M. Kolesik, E. M. Wright, and  J.V. Moloney}
\address{
College of Optical Sciences, University of Arizona, Tucson AZ 85721\\
}

\begin{abstract}
According to a recent experiment, the instantaneous electronic Kerr effect in air
exhibits a strong intensity dependence, the nonlinear refractive index switching
sign and crossing over from a self-focusing to a de-focusing nonlinearity.
A subsequent theoretical work has demonstrated that this has paradigm-changing
consequences for the understanding of filamentation in air, so it is important
to subject the idea of higher-order nonlinearities to stringent tests.
Here we use numerical modeling to propose an experiment capable of
discriminating between the standard and the new intensity-dependent
Kerr-effect models.
\end{abstract}

]

Over the last sesquidecade research into femtosecond filamentation in air has
produced a widely accepted picture in which a dilute plasma, generated
in the high-intensity optical field, is a major determinant of the light pulse
dynamics. For gaseous media, this scenario is based on the notion of a dynamical
balance between the Kerr-induced self-focusing and the counteracting de-focusing
caused by the free electrons~\cite{couairon_femtosecond_2007}. Higher-order nonlinearities
have previously been considered before as an additional
mechanism regularizing the self-focusing collapse. In particular, phenomenological
Kerr-effect saturation at high intensities
(see e.g.~\cite{AkozbekOC01,CouaironPRA03,BergePRA2004} and Sec 3.4. of~\cite{berge_ultrashort_2008})
was explored in filament modeling. In several papers quantum-mechanical
simulations of the electronic polarization response (e.g.~\cite{Nurhuda2008} and references therein)
were used to study the interplay of the saturated Kerr and plasma contributions.
Nevertheless, in the absence of experimental evidence for these effects,
the consensus in the filamentation community was that the standard self-focusing
Kerr effect was sufficient to describe the electronic nonlinearity.
Moreover, many  experiments clearly showed the presence of plasma in filaments,
and this was accepted as the mechanism clamping the intensity~\cite{deng_transverse_2006}.

Recent experimental results by Loriot et al.
\cite{loriot_measurement_2009,loriot_measurement_2010} have the potential
to change this long-standing paradigm. They
found that the instantaneous Kerr nonlinearity exhibits a strong intensity
dependence, and that it not only saturates, but even changes from self-focusing
to de-focusing at large enough intensity.
As a consequence, the plasma is not necessary for the arrest of the self-focusing
collapse, because the Kerr-induced de-focusing at high intensity can alone
be sufficient. Inspired by these findings, the recent modeling
work of Ref.~\cite{Bejot2010} proposed that contrary to our current belief,
ionization-free filamentation is possible in gases.

This has been a surprising development to say the least, and one which
poses many questions. The first one is of course, why the earlier experiments
did not detect such a strong intensity dependence of the Kerr effect? Further,
is it just an accident that the cross-over from self-focusing to de-focusing occurs
in the vicinity of the intensity range where plasma starts to form, or is there
a connection between the two? In particular, central to the interpretation of the
Loriot`s experiment is the assumption that the optical response of the plasma is
isotropic, which is certainly true for the simple Drude model. However, ultrafast birefringence
is know to occur for free electrons in semiconductors due to anisotropic state filling~\cite{Wang2007}, 
and if a similar effect arises for the free-carrier plasma in air it could change the interpretation
of the experiment. And perhaps the most fundamental question is
what is the microscopic origin of the sign-change in the Kerr effect if not due to plasma?

Since the observations of Loriot et al. have such large implications for the
physical picture of filamentation in gases it is important that they be corroborated
in similar experiments, and that the occurrence of higher-order nonlinearities be subjected
to independent experimental testing.
This may not be easy, because accurate quantitative
measurements in filaments are notoriously difficult. It is also conceivable that
the Kerr intensity-dependence could be partially masked by the plasma
induced de-focusing. In fact, this may very well be the reason why this effect has not
been observed earlier. It is therefore the view of the present authors that
this problem needs to be addressed with an utmost caution. Before making conclusions
with regards to the filamentation dynamics, ways should be found to
independently test the interpretation of the Loriot`s experiment.
As a first step in this direction, based on numerical simulations
we propose the design of an experiment which should be capable to discriminate
between the two model candidates: i.e. between the standard Kerr,
and the higher-order Kerr effects.

The proposed experiment is simple in principle. One should use a modest
pulse energy to generate third- and fifth-harmonic radiation, and measure
their respective conversion efficiencies.
To make this feasible, the fundamental frequency should be ``shifted''
to a longer wavelength, perhaps by using an OPA, such
that the third and fifth harmonic energies can be simultaneously measured.
In order to minimise the propagation effects, the focusing geometry should be
tight, resulting in a short interaction volume. The aim is to
compare the fifth-harmonic generation to that of the third harmonic, and
decide if the result is in keeping with the predictions of either of
the competing models. The rationale behind this is the simple fact that
the higher-order Kerr effect should result in a much
stronger harmonic generation. Indeed, in the standard model only the third
harmonic is produced directly from the fundamental, while the fifth (and higher)
harmonics arise via a cascade from the third. On the other hand, the fifth
harmonic can be produced directly from the fundamental in the higher-order
Kerr model.

In our numerical experiments we use the Unidirectional Pulse Propagation Equation
simulator~\cite{KolesikPRE04} which works directly with the real electric field
(as opposed to amplitude envelopes), and therefore naturally and without any modifications
captures all frequency-mixing interactions. We assume the 50~fs input pulse with a central 
wavelength of 1330nm and specify its nominal intensity and beam waist which would be
achieved in a linear focus. The results presented
in this paper were obtained for the nominal focused beam waist of 50~microns,
and the nominal in-focus intensity between $10^{17}$ and $10^{18}$~W/m$^2$.
This tight focusing results in the interaction region only several centimeters
long and is meant to suppress the possible propagation- and plasma-related
effects. For the plasma generation, we use the so-called PPT ionization
rates~\cite{TalebpourOC99} parametrized as in Ref.~\cite{KasparianAPB2000}.
Besides the instantaneous Kerr effect, we include the usual delayed Raman
component~\cite{KolesikPRE04} with the instantaneous-to-delayed ratio of one half.

It should be emphasized that the only difference between the comparative simulation
runs is the different inputs specifying the nonlinear index as a function of
the instantaneous value of the electric field.
For the higher-order Kerr effect model, we have (with suitable electric field units)
$\Delta n = n_2 E^2 +n_4 E^4 +n_6 E^6 +n_8 E^8$ with the
$n_{2k}$ values for air fixed according to~\cite{loriot_measurement_2010}.
The standard model shares the same $n_2$ and has  \protect{$n_4=n_6=n_8=0$}.

\begin{figure}[t]
\centerline{\scalebox{0.5}{\includegraphics[clip=true]{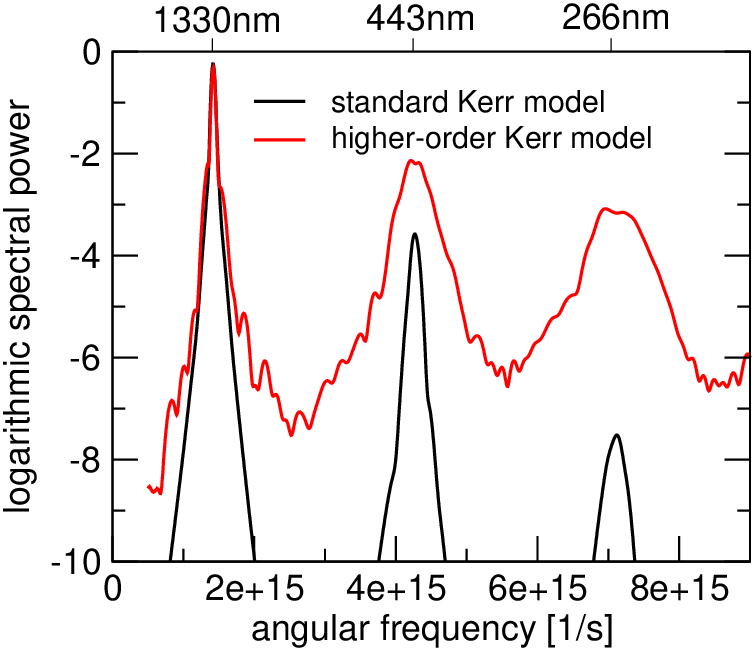}}}
\caption{
\label{fig:spectra}
Angle-integrated spectrum showing third- and fifth-harmonic
generation. The standard and higher-order Kerr effect models
are compared for a tightly focused femtosecond pulse. The
new model predicts stronger third-harmonic generation and
several orders of magnitude more powerful fifth-harmonic radiation.
}
\end{figure}

Let us first look at the typical spectra generated by the two models.
Figure~\ref{fig:spectra} shows the results for the nominal
focal intensity of $10^{18}$W/m$^2$ (the actual intensity reached in
the interaction volume is somewhat smaller). As expected, we can see a
drastic difference. First, in the new model the third-harmonic generation
is almost two orders of magnitude stronger. Second,
the fifth harmonic generation is four to six orders of magnitude
stronger. Thus, if the new model is correct the fifth harmonic should be
easily detected.

\begin{figure}[t]
\centerline{\scalebox{0.5}{\includegraphics[clip=true]{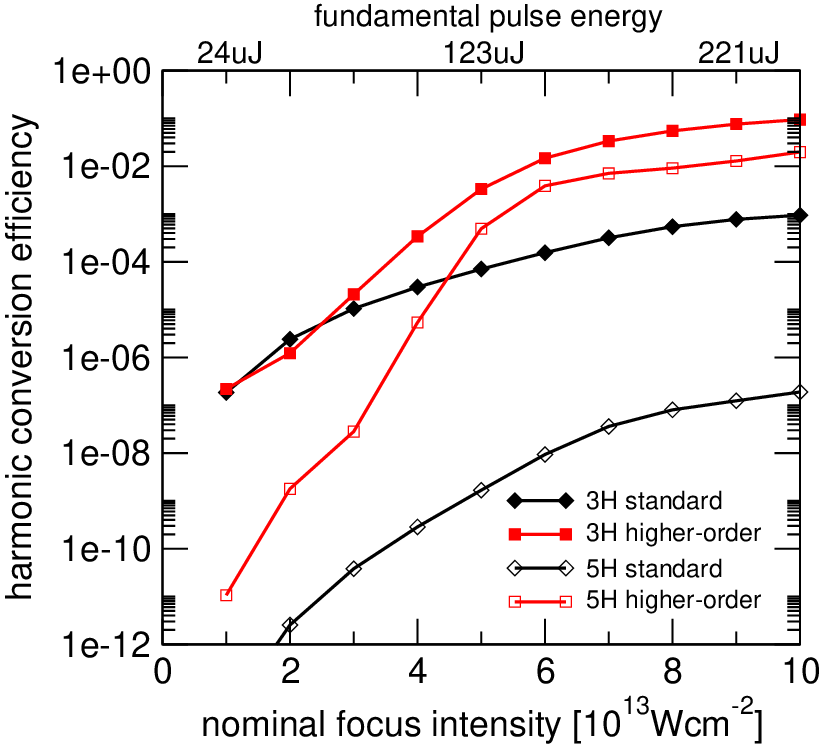}}}
\caption{
\label{fig:H3and5eff}
The harmonic conversion efficiency as a function
of nominal focal intensity/energy.
Full and open symbols represent the third and fifth harmonics,
and squares and diamonds represent the higher-order and
standard model, respectively.
}
\end{figure}

To quantify the above observation further, in Fig.~\ref{fig:H3and5eff}
we show how the third- and fifth-harmonic conversion efficiencies evolve
with the increasing pulse energy (or, equivalently, with the nominal
focal intensity).
The standard model produces the 3H efficiency approaching fraction
of a percent which is in agreement with both experiments
(see e.g~\cite{ThebergeOC05,Xiong08}) and
previous simulations~\cite{kolesik_simulation_2006}.
On the other hand, the 3H efficiency in the
higher-order model can be as high as several percent! To our
best knowledge this has not been observed yet.

However, one could argue that a mere order of magnitude
difference in the third-harmonic conversion efficiency might
be difficult to detect experimentally, because any real experiment
will only probe ``a single model version.'' That is why we now turn to
the fifth harmonic conversion efficiency. Figure~\ref{fig:H3and5eff}
shows that here the difference reaches five orders of magnitude,
and this should indeed be possible to identify in the experiment.
In fact, if the standard model is correct, the fifth harmonic
might be hardly detectable as it should be seven orders
of magnitude weaker than the fundamental. If on the other hand
the new model is correct, the fifth harmonic should be easily detected on the
level of up to a few percent of the fundamental energy.

\begin{figure}[t]
\centerline{\scalebox{0.5}{\includegraphics[clip=tru]{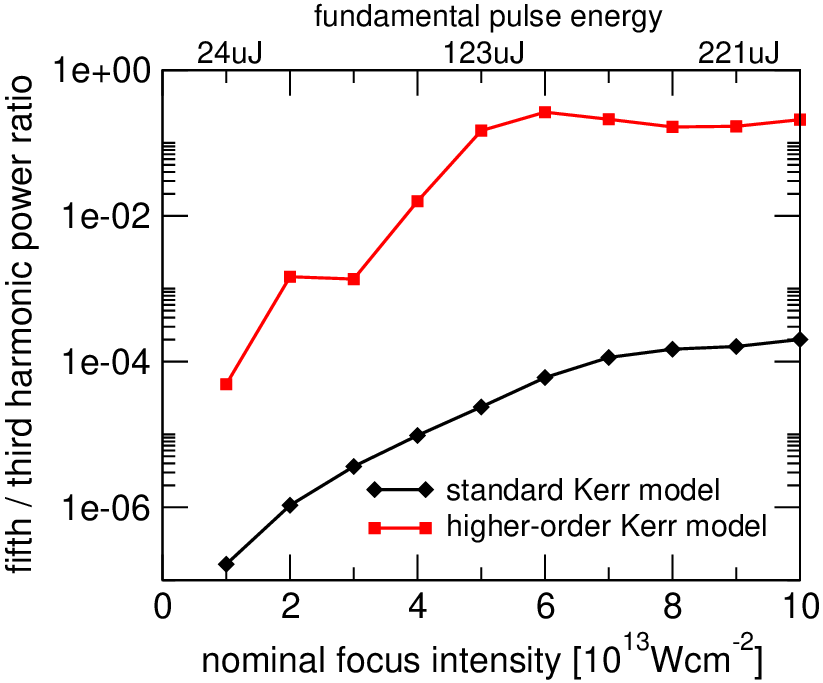}}}
\caption{
\label{fig:H53rat}
Ratio of the fifth- to the third-harmonic energy generated
in the tightly focused femtosecond pulse. The standard model
ratio roughly corresponds to an exponential decrease between
the harmonic orders. In contrast, the new model predicts that
the energy carried by the fifth harmonic can be a mere order
of magnitude less or even comparable to that in the third harmonic.
}
\end{figure}

Another potentially useful telltale sign which an experiment
may look at is the ratio between the fifth and third harmonic
production. This is shown in Fig~\ref{fig:H53rat}. While the
standard model exhibits a small ratio reflecting the fact that
the process is a cascade and the fifth harmonic production
feeds on the third harmonic energy, the two harmonics become
almost comparable in strength in the new model. This occurs
as soon as the focal intensities reach values beyond the
turn-over point of the higher-order Kerr effect. In fact,
it is easy to envision that once the electric field amplitude
increases beyond the turn-over point, the nonlinear index change will
exhibit {\em two} peaks within each half-cycle of the
driving field. This in turn means that the nonlinear index
spectrum contains roughly the same amounts of second and fourth
harmonics giving rise to comparable strengths of the third
and fifth harmonic components in the nonlinear polarization.
In a nutshell, this is the reason why the new higher-order
Kerr model should produce extremely strong fifth-harmonic
radiation.

Let us summarize our findings. Already the present calculations
suggest that the standard model is in a better agreement with
previous experiments on the third-harmonic generation. It exhibits
the conversion efficiencies comparable with the
measured values. In contrast, the new model predicts
significantly stronger third harmonic, with its energy reaching
up to several percent of the fundamental. To our knowledge such a strong
third-harmonic generation has not been observed. However, we should keep in mind that due to
the model parameter uncertainties the possibility can not be excluded
that the simulated efficiency value may suffer from
a systematic deviation.

That is why we propose an experiment which should compare
the fifth and third harmonic generation. We have shown that
if the higher-order Kerr effect model is correct, the experiment
will detect significant fifth-harmonic generation. Moreover,
at higher pulse energies
the relative strength of the fifth to the third harmonic should
reach values of the order of $10^{-1}$. On the other hand,
if the standard model is correct, this ratio should be about
four to five orders smaller. This is a significant qualitative difference,
and it should be obvious from the experimental results which
of the competing models should be preferred.

In our view, the importance of the independent corroboration of
the experiment by Loriot et al. can not be overemphasized. Their
results imply literally a change of paradigm in our understanding
of femtosecond filamentation, put in doubt concepts such as
plasma-mediated intensity clamping, and essentially invalidate
the very model of femtosecond filamentation that has been
used for over a decade. It is thus our aim and hope that this work
will motivate and inspire experimental efforts to solve this exciting
and important problem.

This work was supported by Air Force Office of Scientific Research
under contract FA9550-07-1-0010.

%\bibliographystyle{ol}
%\bibliographystyle{osajnl}
%\bibliography{../filamentation}

\end{document}